# Patent Classifications as Indicators of Intellectual Organization

Loet Leydesdorff

Amsterdam School of Communications Research (ASCoR), University of Amsterdam

Kloveniersburgwal 48, 1012 CX  Amsterdam, The Netherlands

loet@leydesdorff.net ; http://www.leydesdorff.net

**Abstract**

Using the 138,751 patents filed in 2006 under the Patent Cooperation Treaty, co-classification analysis is pursued on the basis of three- and four-digit codes in the International Patent Classification (IPC, 8th edition). The co-classifications among the patents enable us to analyze and visualize the relations among technologies at different levels of aggregation. The hypothesis that classifications might be considered as the organizers of patents into classes, and that therefore co-classification patterns—more than co-citation patterns—might be useful for mapping, is not corroborated. The classifications hang weakly together, even at the four-digit level; at the country level, more specificity can be made visible. However, countries are not the appropriate units of analysis because patent portfolios are largely similar in many advanced countries in terms of the classes attributed. Instead of classes, one may wish to explore the mapping of title words as a better approach to visualize the intellectual organization of patents.

## 1. Introduction

Soon after his original publication in *Science* about using citation indexing in scientific literature (Garfield, 1955), Garfield (1957) published a less-known paper in the *Journal of the Patent Office Society* entitled "Breaking the subject index barrier—a citation index for chemical patents." Based on Shephard's citation index in law (Adair, 1955), Garfield generalized the notion of citation indexing as a tool for mapping an associative network as distinct from a hierarchically organized index. As he stated (at p. 472):

> Keeping the user in mind, the conscientious indexer will translate the terminology and phraseology of the author into a standardized and more usable form. However, the indexer is, of necessity, primarily guided by the subject content which authors provide.
>
> The indexer also faced with a practical economic barrier cannot index with the almost infinite depth to be found in the Citation Index. The Citation Index breaks this "barrier" by presenting subject matter in bibliographical arrays which are neither alphabetical nor classified but associative.

Unlike the imposed categories of classificatory indices, associations by citation are generated by the authors and inventors themselves. The network structure *emerges* as a property from the aggregates of individual actions. Since this network is dynamic, it can be expected to develop self-organizing properties. Order—such as the grouping into disciplines and specialties—is the result of interaction between top-down and bottom-up dynamics (Kauffman 1993; Cilliers 1998).

Although the *self-organizing* dynamics of the communication can be expected to prevail in the case of scientific literature (Kuhn, 1962; Price, 1965; Luhmann, 1990; Leydesdorff, 1995), the situation for patents is very different. Patents are regulated by law; legislation is (predominantly) *organized* at the level of national states. Patent citations and classifications are often added by the patent examiners. Can patent citations be used to map the underlying science base of patents (Jaffe & Trajtenberg, 2002; Leydesdorff, 2004; Porter & Cunningham, 2005; Sampat, 2006)?

Citations are added to the front pages of the patents by the examiners to varying extents (Cockburn *et al*., 2002), and the procedures differ among national, regional, and worldwide operating offices (Criscuolo, 2004). Patent citations can also be expected to fulfill other functions such as protecting an industrial portfolio, etc. (Mogee & Kolar, 1999). Because of these legal and economic functions, patent citations have been considered from perspectives different from the mapping of their knowledge base. For example, patent citations have been used for measuring the economic value of patents (Trajtenberg, 1990; Hall *et al*., 2002; Sapsalis *et al*., 2006). This has additionally been done at the level of innovation systems or even companies (Engelsman & Van Raan, 1993, 1994; Breschi *et al*., 2003; Leten *et al*., 2007). In summary, the different contexts of innovations (the market value, the legal status, and the knowledge base) provide patent citations with a variety of meanings.

In order to focus on the science base of patents, some scholars have analyzed the so-called "non-patent literary references" (NPLR) among the patent citations (Narin &

Olivastro, 1988 and 1992; Grupp & Schmoch, 1999; Leydesdorff, 2004). In a recent study, Leydesdorff & Zhou (2007) examined the disciplinary background of NPLR using the journals being cited in the case of nanotechnology patents. We found in these case studies that NPLR tend to be dominated by references to general science and professional journals. These journals cannot easily be identified with specific journal categories or disciplines (Guan & He, forthcoming).

In other words, contrary to Garfield's (1957) idea, we may have to pay more attention to the indexing and the expected indexer effects (Healey *et al*., 1986). In the case of patents, indexing is part of a search process by the examiner which may lead to the addition of citations, but in any case to the careful selection of primary and secondary categories for the disclosure. The primary classes may be seen as defined areas of interest; the class title gives an indication of the content of the class (WIPO, 2006, at p. 10). The secondary classes can be viewed either as exploratory classifications (Verspagen, 2005) or as reinforcing the primary classification (Breschi *et al*., 2000). The subclass title indicates as precisely as possible the content of the subclass (WIPO, 2006, at p. 10). The examiners thus add intellectual content to the patent structure (Larkey, 1999). The two processes of application and evaluation are more interwoven than in the science system. Furthermore, one would expect different networks to emerge from the different constraints of procedures in the cases of national, regional, and worldwide applications.

## 2. Various patent databases

The database of the U.S. Patent and Trade Office (USPTO) contains all the data since 1790. Patents are retrievable from this date as image files, and after 1976 also as full text. The html-format allows us to study them in considerable detail (Leydesdorff, 2004). The European Patent Office (EPO) was established as a transnational patent office in 1973. This database is also online, but in a less accessible pdf-format. Furthermore, 135 nations (as of December 31, 2006) were signatories of the Patent Cooperation Treaty (PCT) of 1970, which mandated the World Intellectual Property Organization (WIPO) in Geneva to administer fee-based services (since 1978). This database is well organized and in the html-format. The various services enable users in member countries to file international applications for patents. Like the European Union, several world regions have established regional patent offices.

The various offices provide applicants with a number of choices which imply different procedures and timelines. For example, the USPTO operates on the basis of "first to invent," while the EPO uses "first to file" as a criterion. If the applicant wishes to protect an invention in countries outside the country of its origin, s/he can file for a patent in each country in which protection is desired, or to a regional office (e.g., EPO), or file an international application under the Patent Cooperation Treaty procedure (OECD, 2005: 54 ff.). Various factors (e.g., the costs of patenting, the time taken to grant patents, differences in rules regarding the scope of patents, etc.) influence the decision on whether to follow one procedure or another.

This variety of (partially overlapping) procedures complicates the use of patent statistics. In the USPTO the inventor and his/her attorney are obliged to provide a list of references describing the state of the art—the so-called "duty of candor" (Michel & Bettels, 2001). EPO examiners, and not inventors or applicants, add the large majority of patent citations (Criscuolo & Verspagen, 2005, at p. 10). While in the USPTO applications are examined automatically, the EPO considers an application as a request for a "patentability search report." These reports contain citations from patents and non-patent documents that have either been suggested by the inventor or added by the patent examiner (Criscuolo, 2004, at pp. 92f.). Applications at the WIPO for intellectual property protection under the regime of the PCT protocol require an International Search Report (ISR) and a written opinion by the examiner about the patentability of the invention (OECD, 2005, at p. 57). Although the initial investigations are usually carried out by the receiving offices, the international extension can be expected to lead to a further streamlining of the patent citations with reference to their economic value and legal protection against possible litigation in court.

From the perspective of information science and technology, patent classifications provide us with the outcomes of major investments of the patent offices to organize the patents intellectually. The International Patent Classification system (IPC8) is currently (since January 1, 2006) in its eighth edition, using a 12-digit code containing 70,000 categories (WIPO, 2006 and 2007). The European classification system (ECLA) builds on this IPC and extends the number of categories to 134,000. The USPTO has its own classification system, which has been used extensively in economic research (Jaffe & Trajtenberg, 2002). This classification scheme currently employs up to 430 classes and 140,000 subclasses.

Attempts to map patents for the purpose of analyzing economic activities in terms of the technologies involved have been moderately successful. The OECD has for this purpose defined "triadic patent families" which counteract upon "home advantage effects" in the various databases (Criscuolo, 2006). A patent is a member of a patent family *if and only if* it is filed at the European Patent Office (EPO), the Japanese Patent Office (JPO), and is granted by the US Patent & Trademark Office (USPTO) (Eurostat, 2006). However, this *institutional* integration of the different databases into a single set of files that can then also be used for normalization has hitherto remained problematic. The recent integration of the various databases (OECD, WIPO, EPO, etc.) into the framework of PATSTAT cannot be expected to resolve the problem because this relational database is developed at the institutional level (Magnani & Montesi, 2007).

Several research teams have invested in generating concordance tables between the International Standard Industrial Classification (ISIC) and the International Patent Classifications (IPC) (Evenson & Puttnam, 1988; Verspagen *et al.*, 1994; Verspagen, 1997). In a recent validation study, Schmoch *et al.* (2003, at p. 58) concluded that the correlations between these tables are low. The authors cite Grupp *et al.* (1996, at p. 272) that "industries do not represent homogeneous technologies." From the perspective of industrial economics, the IPC itself is mostly taken as a given because patents are considered as input indicators. The purpose of this study is to open this black box and to consider patents as outputs of the R&D system. How do the classifications map the intellectual organization of patents?

### 3. Classifications

While scientific publications are organized in terms of journals, the primary system for organizing patents intellectually is classifications. In a recent review of methodologies in patent statistics, Dibiaggio & Nesta (2005) argued that technology classes could be the most appropriate unit of analysis for exploiting the information contained in the patent databases. Co-classification analysis has been explored in the study of scientific literature, but in scientometrics co-classification analysis has been less successful than co-citation analysis because the classifications are imposed, while the citations are not (Todorov, 1988; Tijssen, 1992b; Spasser, 1997). From a methodological perspective, co-occurrence data (co-words, co-classifications, co-citations, etc.) can equally be used for the mapping (Leydesdorff, 1987; Tijssen, 1992a; Leydesdorff & Vaughan, 2006).

In other words, it seems worthwhile to investigate whether co-classification analysis at the level of the database can provide us with an angle for the mapping of the intellectual organization of the patents. From the perspective of information analysis, the PCT database of the World Intellectual Property Organization in Geneva has the advantage that all records are gathered according to a common standard. Furthermore, one can expect that mainly patents of a certain economic and technological value will be extended for protection beyond the domestic market. Thus, mapping these patents can be expected to show the fields of technological specialization of each country. The disadvantage remains that patents under the PCT regime are a specific subset of the total set of patents in the world. Some countries may use this route more than others. However, the PCT procedure is increasingly used for patent applications. The number of applications increased from around 24,000 in 1991 to 110,000 in 2002 (OECD, 2005, at p. 7). Currently, more than 135,000 applications are registered yearly (WIPO, 2007).

From the perspective of my research question, the problem that the PCT set is a specific subset is ameliorated by the high level of codification within this set because of the intensive development of the IPC by the WIPO. If one were unable to retrieve structure in this relatively well-organized set, then the more fuzzy sets would be even more difficult to analyze. The national, regional, and international procedures for an application under the PCT regime take approximately 30 months, but after this period the patent is designated for all the countries indicated (Figure 1). This delay is sometimes considered as an advantage because it provides the applicant with more time to decide whether or not to seek a national or regional patent.

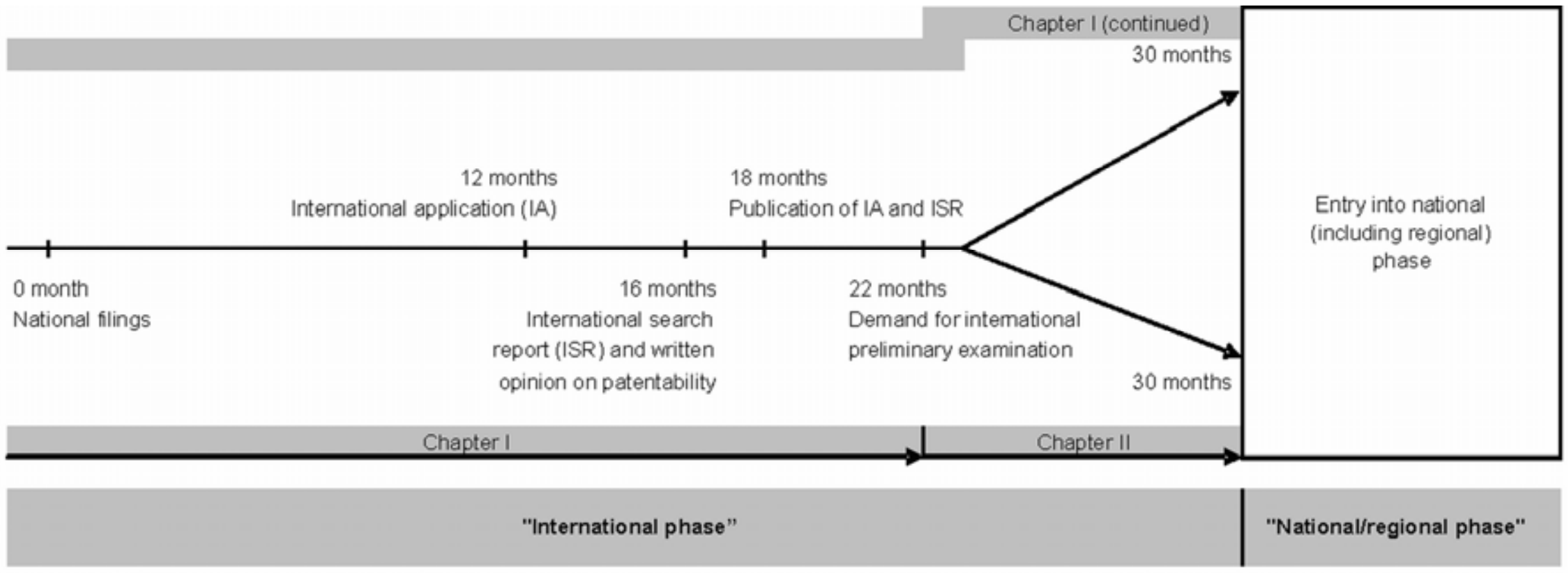

**Figure 1**: Timeline for PCT procedures (Source: OECD, 2005, at p. 57)

Patent co-classifications were already mentioned in the *OECD Manual* of 1994 as a potential indicator of linkages among technologies (OECD, 1994, at p. 52). However, the emphasis in the literature has been on patent citations because of (1) the analogy with citations in the scientific literature, and (2) the interest in patent citations as indicators of economic value (Breschi & Lissoni, 2004). Hall *et al*. (2002) grouped the classes of the U.S. classification into six technological categories and 36 subcategories, but this research was not used intensively in further research for mapping the knowledge structure of patents. Breschi *et al*. (2002, 2003) have explored the mapping of firm portfolios and their technological coherence in terms of patent classifications. Using the cosine and other measures of similarity, these authors concluded that relevant measures of the technological proximity of the classes can be retrieved from the database (Ejermo, 2005).

## 4. Methods and materials

### *4.1. Data*

Because the WIPO database of the PCT applications is fully accessible online, has worldwide coverage, and is carefully indexed using the latest version of the IPC, I decided to download one year of data, that is, the 2006 data, from this database. The downloads were done in the second week of January 2007. The dataset was fixed at that date at 138,741 patents with a publication date in 2006. Actually, 138,751 patents were retrieved, but the difference of ten patents is negligible given the large numbers.

It was decided to use publication dates instead of application dates because the applications are brought online in a moving process. Thus, the number of patents using application dates as the search code varies from day to day. For example, on 18 February 2007, 73,506 patents with application dates in 2006 were available, while this number increased to 76,237 one week later, on February 25. Even the number of patents with application dates in 2005 changed in this week from 129,841 to 130,066.

The patents were downloaded and brought under the control of relational database management using dedicated software routines. Table 1 provides the descriptive statistics.

| | *N* | *N / patent* | |
|---|---|---|---|
| *patents* | 138,751 | | |
| *inventors* | 365,699 | 2.63 | |
| *applicants* | 473,367 | 3.41 | |
| *classifications* | 325,393 | 2.35 | |
| *designations* | 13,847,717 | 99.80 | } 102.8 |
| *regional* | 415,729 | 3.0 | |
| *countries* | 225 | includes regions | |

**Table 1**: descriptive statistics of the data

Using the addresses of the inventors for the attributions, the distributions of inventors and applicants over various countries are shown in Figure 2. These distributions exhibit the well-known logarithmic shape of a Lotka-distribution. The fit is almost perfect ($r > 0.99$) for the inventors.

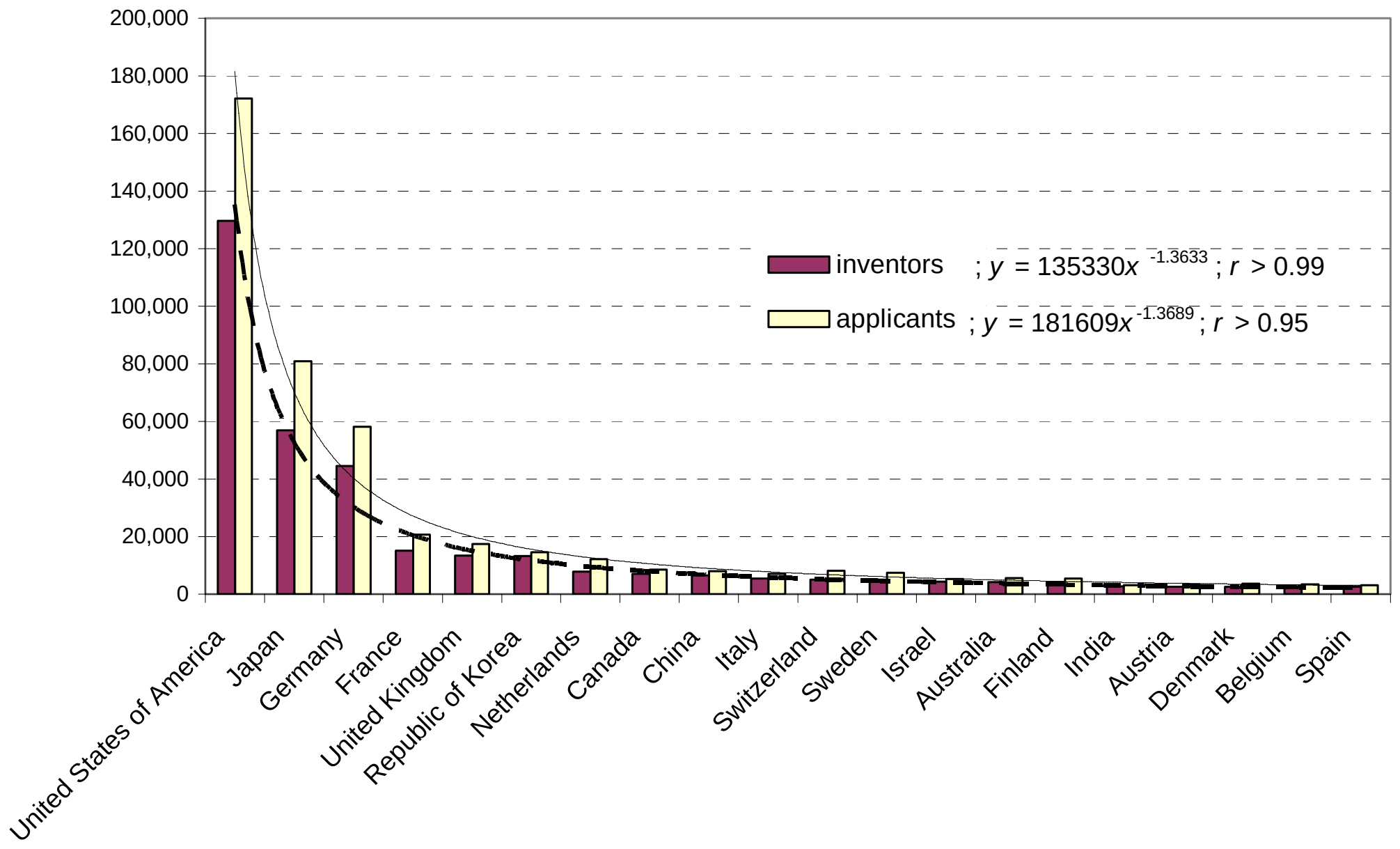

**Figure 2**: Distribution over major patenting countries ($N$ patents > 1000)

Relatively small countries like Korea and the Netherlands are more important contributors to the database than Italy and China. In larger countries, domestic patenting may play a more important role than in smaller ones. Within the EU, European patents increasingly replace domestic patenting. For example, only 2,152 of the 17,095 patents with a German inventor are patented in Germany itself. Note that Russia is not a major player in this system.

As summarized in Table 1, each patent has on average 2.6 inventors and 3.4 applicants. However, 334,737 inventors (> 99%) are also co-applicants; only 138,630 applicants are non-inventors. This is approximately one per patent. Figure 2 shows that the practices of co-application by inventors vary among countries. In the case of South Korea, for example, mostly inventors seem to apply.

The number of classifications per patent is on average 2.4. The number of designations is of the order of 100. Further analysis of these co-designations may be interesting from the perspective of industrial strategies and spillovers. As noted, the classifications are very detailed, using up to 12 digits. The main classes are contained in a four-“digit” categorization (WIPO, 2006). Table 2 provides class A01 and its four-digit extensions as an example. In the 2006 data, 121 main categories (at the three-digit level) were included with elaboration into 623 categories at the four-digit level.

| A01 | AGRICULTURE; FORESTRY; ANIMAL HUSBANDRY; HUNTING; TRAPPING; FISHING |
|---|---|
| A01B | Soil working in agriculture or forestry; Parts, details, or accessories... |
| A01C | Planting; Sowing; Fertilising |
| A01D | Harvesting; Mowing |
| A01F | Processing of harvested produce; Hay or straw presses; Devices for... |
| A01G | Horticulture; Cultivation of vegetables, flowers, rice, fruit, vines,... |
| A01H | New plants or processes for obtaining them; Plant reproduction by... |
| A01J | Manufacture of dairy products |
| A01K | Animal husbandry; Care of birds, fishes, insects; Fishing; Rearing or... |
| A01L | Shoeing of animals |
| A01M | Catching, trapping or scaring of animals; Apparatus for the destruction... |
| A01N | Preservation of bodies of humans or animals or plants or parts thereof;... |

**Table 2**: The first category (“A01”) of the IPC with its sub-classifications as an example

From the perspective of visualization, the 121 categories at the three-digit level are optimal, since the screen becomes unreadable with larger numbers. The user could then be enabled to zoom into the four-digit level. However, in such a hierarchical approach one would lose the lateral links provided by the co-classification analysis. In this study, I first focus on the network of co-classifications at the three-digit level and subsequently at the four-digit level, and then analyze the level of detail available at the level of the nations participating in the database. The analytical insights may help us to understand which approach is more feasible and meaningful.

One four-digit category was added to the patents by hand, using the search facility of the European Patent Office at http://ep.espacenet.com/advancedSearch?locale=en_EP. The EPO recently made a substantial investment by developing the code "Y01N" as an additional tag to the existing database for the nano-categories (Scheu *et al*., 2006; Hullmann, 2006).[1] The tag is relevant because of the interest in policy circles in the evaluation of the current efforts to stimulate nanotechnology (Braun & Meyer, 2007). Since the EPO database can also be searched for PCT applications, 762 records could be matched with this tag.[2] Thus, at the four-digit level, I work with 624 categories.

### *4.2. Methods*

Both at the three-digit and the four-digit level, a matrix was constructed with the patents as the units of analysis and the classification codes as column variables. The analysis focuses on the relations among the variables. For that purpose, I use factor analysis (Varimax Rotation in SPSS) and visualization techniques from social network analysis after normalization of the variables using the cosine. The factor analysis informs us about the structure in the matrix, while the cosine-normalized matrices allow for the visualization and the study of centrality measures (Leydesdorff, 2007). Whenever a submatrix is extracted—for example, for a country on the criterion of the institutional address of the inventor—these same analytical techniques are used.

---

[1] The class is further subdivided into Y01N2 for Bio-nanotechnology, Y01N4 Nanotechnology for information processing, storage and transmission, Y01N6 Nanotechnlogy for materials and surface science, Y01N8 Nanotechnlogy for interacting, sensing or actuating, and Y01N10 Nanooptics.

[2] In the previous organization the field "B82B: Nano-structures: Manufacture and treatment thereof" corresponded to the special class CL/977 which was added to the USPTO. This category matched only 275 patents in 2006 (Leydesdorff & Zhou, 2007).

Of the 138,751 patents, only 135,536 patents contained valid classifications. As noted, the number of classifications in this data was 121 and 624 at the level of three and four digits, respectively. From these matrices, subsets were extracted for 126 countries. Additionally, an aggregated file of countries versus classification codes was generated for further analysis. The latter file enables us to study the distributions of classifications over countries or, *vice versa*, the distributions of countries over classifications. For example, one can raise the question how patents with the IPC-classification "B82B: Nano-structures: Manufacture and treatment thereof" or the ECLA-tag for nanotechnology "Y01N" are distributed across the countries.

The two basic matrices of patents versus classification codes are extremely sparse: in the case of three-digit classes only 1.14% of the cells have a non-zero value and in the four-digit case only 0.25%. Because of the large numbers of zeros the cosine between the variable vectors is a better measure of similarity than the Pearson correlation coefficient (Ahlgren *et al*., 2003). Unlike the latter, the cosine normalizes with reference not to the arithmetic mean, but to the geometric mean. Using the cosine values, one can span a vector space which can be used for visualization purposes (Salton & McGill, 1983; Breschi *et al*., 2003).

The freeware program Pajek (available for academic purposes at http://vlado.fmf.uni-lj.si/pub/networks/pajek/) is used for the visualizations of the cosine-normalized patterns; UCINet and SPSS are used for the statistical analysis. For each country, a cosine-normalized input-file for Pajek is brought online at http://www.leydesdorff.net/wipo06/index.htm. These files allow users to freely choose a visualization routine for a specific country, since Pajek files can be exported into other formats. I provide examples below using the available visualization routines with the purpose of conveying the analytical argument. The algorithm of Furchterman & Reingold (1991) is used in this version for most of the visualizations because—unlike using Kamada & Kawai (1989)—the isolates remain meaningfully visible using this algorithm.

Within the visualizations, the size of the nodes is set proportionally to the *logarithm* of the number of patents in the category. The lines are made proportional to the cosine values in five equal steps of 0.2. A threshold is set at the cosine ≥ 0.05 in order to weed out incidental variation. In most cases, greyshades are added using the *k*-core algorithm for the partitioning in order to support the readability of the visualizations.

## 5. Results

### *5.1. The 3-digit level*

Let us first inspect the overall picture for the 121 patent classifications at the three-digit level and the 135,536 patents. Figure 3 shows that the set is *not* well connected. Fifty-six of the 121 categories are not connected to others at the level of cosine ≥ 0.05; only 13 are more strongly connected (as *k*-cores); 52 classes are connected in weak graphs. I added labels to the 13 categories which form more strongly connected *k*-cores. Upon visual inspection, it seems that the well-connected sets represent chemical industries and biotechnological applications in agriculture.

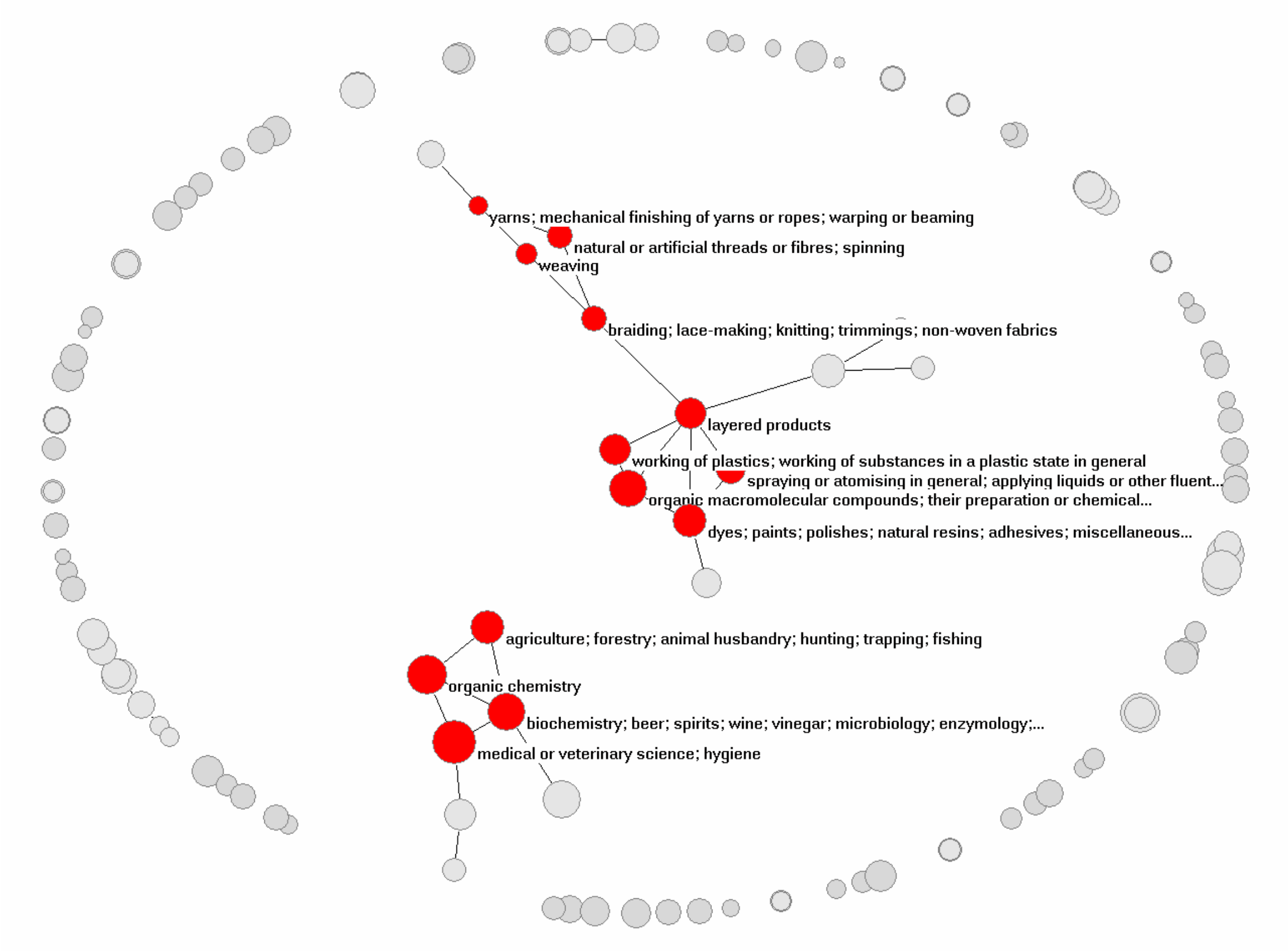

**Figure 3**: 121 patent classifications at the 3-digit level; $N = 135{,}536$; cosine ≥ 0.05; 2D-visualization based on the algorithm of Fruchterman & Reingold (1991).

In other words, the co-classifications do not reveal a clear structure. Unlike scientific citations, one would not expect to find the operation in this data of structure-generating mechanisms like the Matthew effect (Merton, 1968) or preferential attachment (Barabási, 2002; cf. Leydesdorff & Bensman, 2006). On the contrary, one would expect the index to enable the patent offices to distribute the patents over categories. The number of patents per class is intentionally kept down, but a new subclass can be formed to accommodate overflow (Larkey, 1999). Notice also that in the analysis, all PCT applications are considered for a certain time period; this would equal to an analysis of all scientific publications for a given time period. Would one observe a lot of structure within such an exercise?

A next question is whether countries show specific profiles within the larger set. I explore this below using Germany and China as examples. Germany is one of the largest share-holders in the PCT applications, while China is at the ninth position (Figure 2). Germany, of course, has a mature industrial structure, while the Chinese system has been booming during the last decade or so. The German patent set of 17,095 patents covers all 121 patent categories (Figure 4); the Chinese one based on 3,084 patents contains 109 of these categories (Figure 5). Table 3 compares the two sets with the global sets in terms of network statistics.

| three digits;<br>*cosine* > 0.05 | *Global set*<br>(*N* = 121) | *Germany*<br>(*N* = 121) | *China*<br>(*N* = 109) |
|---|---|---|---|
| *Density* | 0.008 | 0.008 | 0.020 |
| *% Degree centralization* | 4.31 | 2.56 | 9.33 |
| *% Closeness centralization*[3] | n.a. | n.a. | n.a. |
| *% Betweenness centralization* | 0.78 | 0.90 | 21.28 |
| *Clustering coefficient* | 0.198 | 0.089 | 0.233 |

**Table 3**: Network statistics of the cosine-normalized matrices for the German, Chinese, and global sets of patents classified at the three-digit level.

[3] Closeness centralization cannot be computed since the networks are not weakly connected.

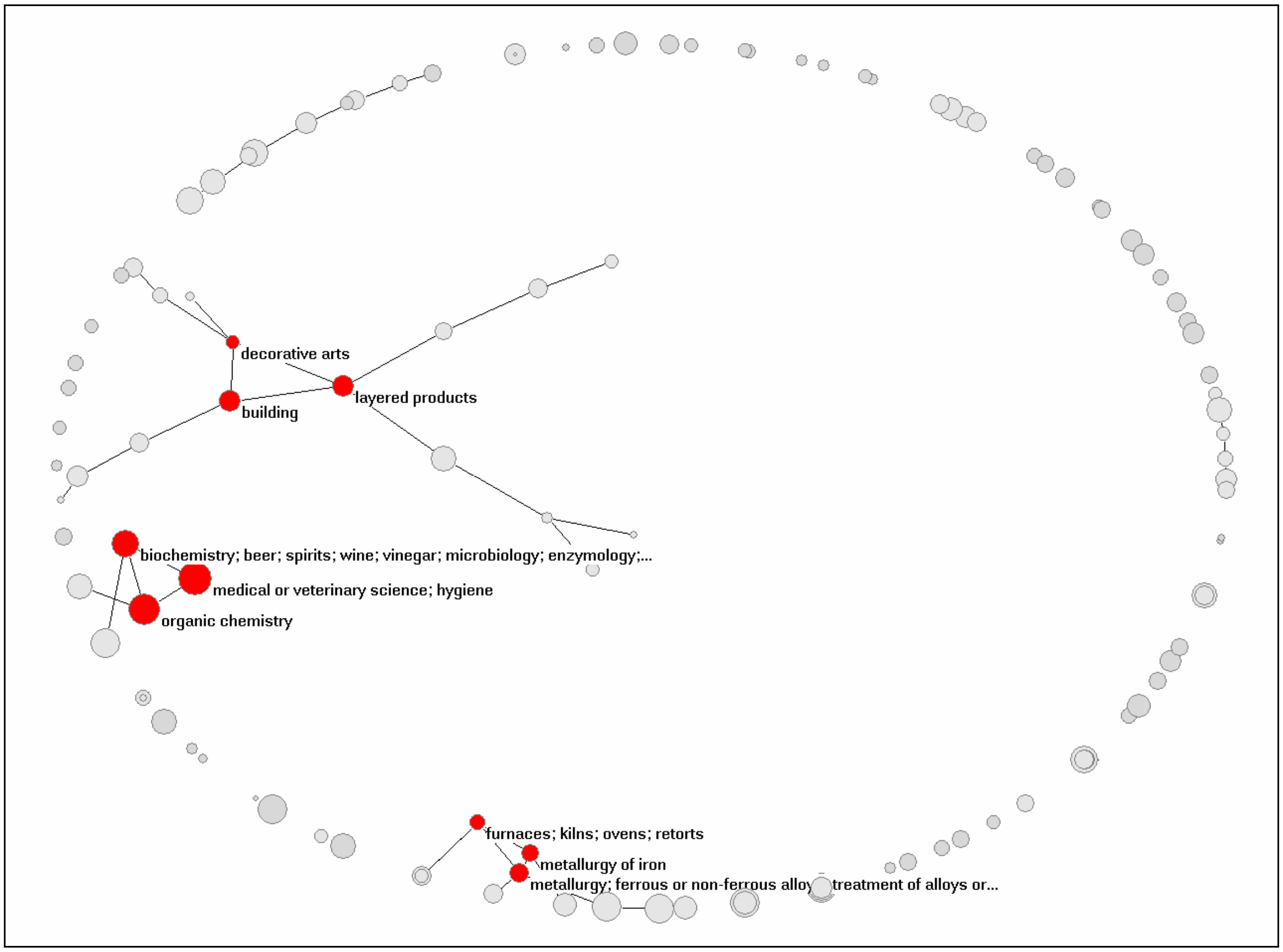

**Figure 4**: Co-classifications of 17,095 of German patents at the 3-digit level; cosine ≥ 0.05; 2D-visualization based on the algorithm of Fruchterman & Reingold (1991).

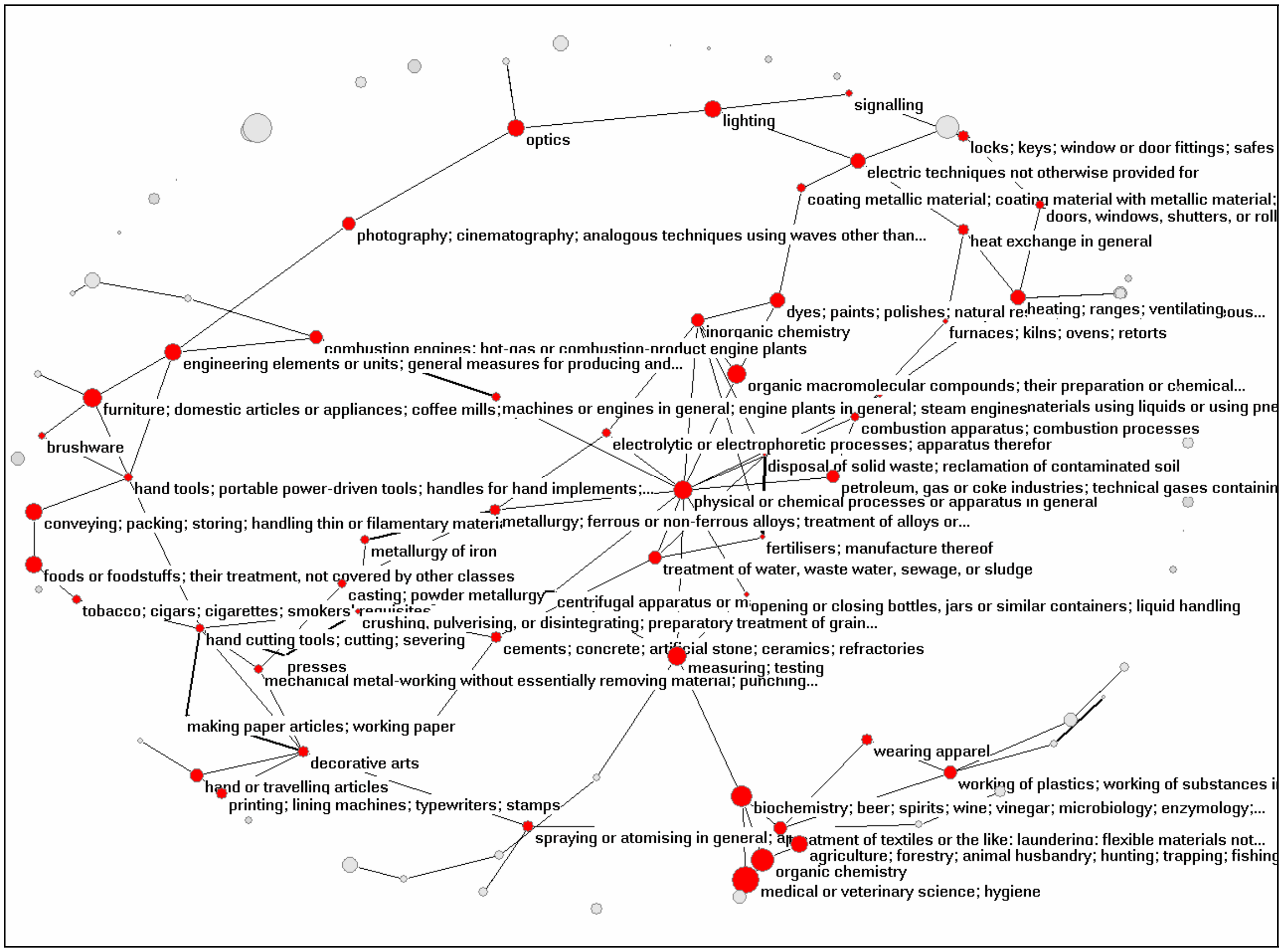

**Figure 5**: Co-classifications of 3,084 Chinese patents at the 3-digit level; cosine ≥ 0.05; 2D-visualization based on the algorithm of Fruchterman & Reingold (1991).

The structure in the German set is of the same order as for the complete set, while there is much more structure in the Chinese set. In Figure 5, 82 of the 109 classifications are connected into a weak component. All three sets (the global one and the two for Germany and China, respectively) have in common a core group of "medical or veterinary science; hygiene," biochemistry, and organic chemistry.

Factor analysis of the underlying matrices of patents versus classes confirms that there are no pronounced eigenvectors: more than 55 eigenvectors explain more than an average variable; none of the eigenvectors explains more than 2% of the common variance (Leydesdorff & Hellsten, 2005). The Chinese network is a bit more pronounced than the German one or the one at the global level.[4] In other words, the networks are very flat and the categories are not obviously informative.

The lack of organization in the data suggests taking a closer look at betweenness centrality as another measure for connectedness and coherence in the profiles (Freeman, 1997; Breschi *et al*., 2003; Leydesdorff, 2007). Table 4 provides the top-25 categories in terms of the percentage of betweenness centrality for the two countries.[5]

| Germany | % | China | % |
|---|---|---|---|
| 1. medical or veterinary science; hygiene | 7.5 | 1. measuring; testing | 13.2 |
| 2. engineering elements or units; general measures for producing and... | 6.9 | 2. medical or veterinary science; hygiene | 13.1 |
| 3. measuring; testing | 6.8 | 3. furniture; domestic articles or appliances; coffee mills; spice mills;... | 10.1 |
| 4. physical or chemical processes or apparatus in general | 5.9 | 4. physical or chemical processes or apparatus in general | 8.9 |
| 5. vehicles in general | 5.8 | 5. basic electric elements | 8.5 |
| 6. dyes; paints; polishes; natural resins; adhesives; miscellaneous... | 3.5 | 6. engineering elements or units; general measures for producing and... | 7.9 |
| 7. working of plastics; working of substances in a plastic state in general | 3.5 | 7. organic macromolecular compounds; their preparation or chemical... | 5.7 |
| 8. basic electric elements | 3.3 | 8. computing; calculating; counting | 5.4 |
| 9. furniture; domestic articles or appliances; coffee mills; spice mills;... | 2.7 | 9. layered products | 4.8 |
| 10. layered products | 2.6 | 10. vehicles in general | 4.6 |
| 11. conveying; packing; storing; handling thin or filamentary material | 2.5 | 11. foods or foodstuffs; their treatment, not covered by other classes | 3.8 |
| 12. organic macromolecular compounds; their preparation or chemical... | 2.3 | 12. conveying; packing; storing; handling thin or filamentary material | 3.8 |
| 13. machine tools; metal-working not otherwise provided for | 1.7 | 13. agriculture; forestry; animal husbandry; hunting; trapping; fishing | 3.0 |
| 14. ammunition; blasting | 1.7 | 14. cements; concrete; artificial stone; ceramics; refractories | 2.7 |
| 15. coating metallic material; coating material with metallic material;... | 1.6 | 15. treatment of textiles or the like; laundering; flexible materials not... | 2.7 |
| 16. electric techniques not otherwise provided for | 1.5 | 16. electric techniques not otherwise provided for | 2.7 |
| 17. building | 1.3 | 17. electric communication technique | 2.4 |

[4] The first eigenvector explains 1.7% of the common variance in the Chinese case, versus 1.2% for both the German and the total set.
[5] The betweenness centrality is calculated from the cosine-normalized matrix, but before a threshold is set. If the matrix is not normalized, betweenness centrality is often overshadowed by degree centrality, since a "star" in a network is also "between" many nodes (Leydesdorff, forthcoming).

| | | | | | |
|---|---|---|---|---|---|
| 18. | agriculture; forestry; animal husbandry; hunting; trapping; fishing | 1.3 | 18. | dyes; paints; polishes; natural resins; adhesives; miscellaneous... | 2.3 |
| 19. | spraying or atomising in general; applying liquids or other fluent... | 1.2 | 19. | hand cutting tools; cutting; severing | 2.3 |
| 20. | optics | 1.1 | 20. | optics | 2.2 |

**Table 4**: top 20 classes at the three-digit level and the percentages of betweenness centrality for Germany and China, respectively.

These results confirm the impression that the Chinese system is more integrated than the German one in terms of these measures. Although one can observe differences in the ranking, these differences are not obviously informative. The similarities are also considerable: the two tables have fourteen of the twenty categories in common.

### *5.2 The 4-digit level*

At the 4-digit level, the lack of structure is less obvious, but still considerable at the level of the aggregated set. 115 categories are not connected at the 0.05 level for the cosine. There are a few dense clusters in traditional industries (fertilizers, chemistry, etc.).

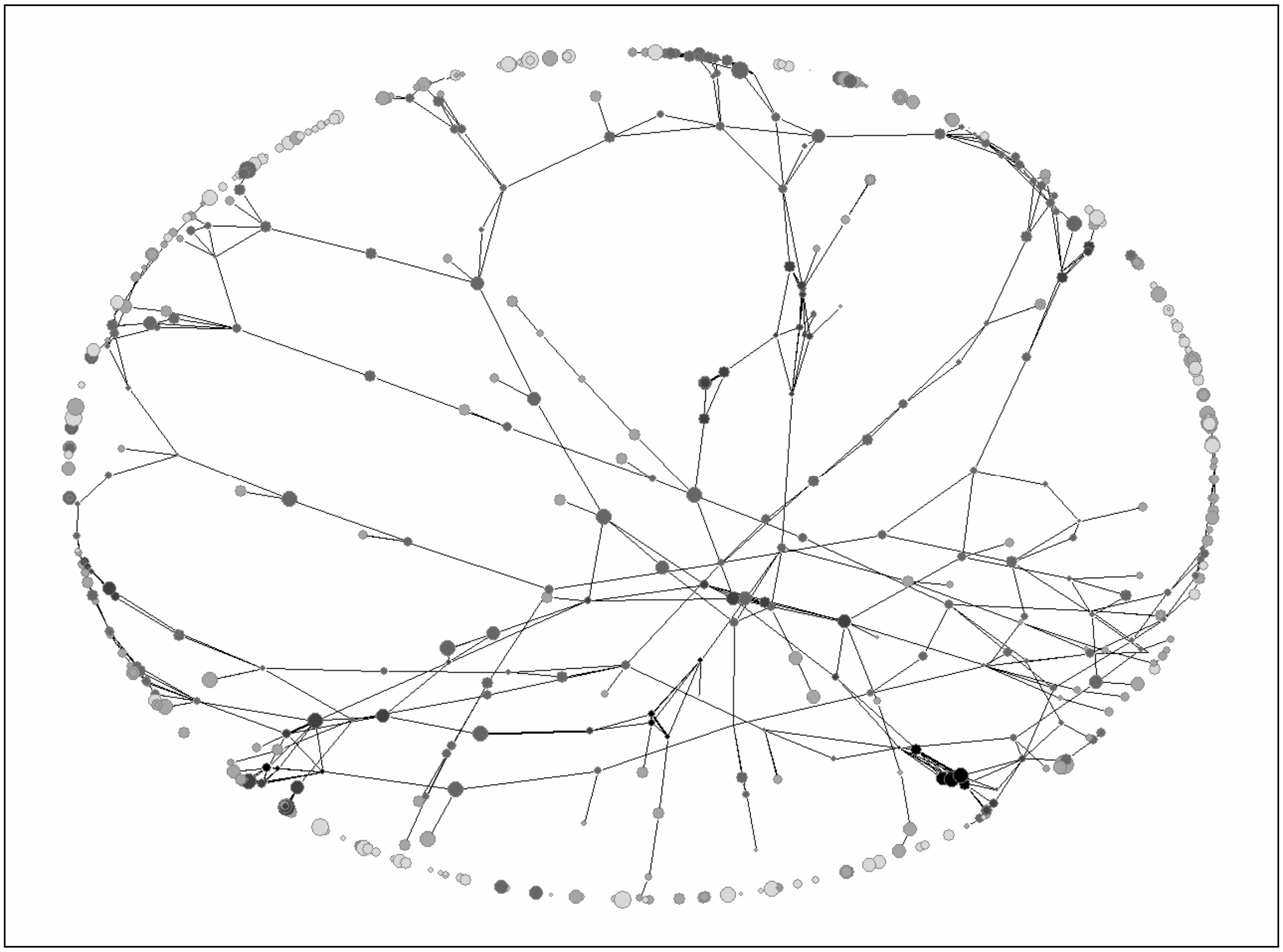

**Figure 6**: 624 patent categories versus 135,536 patents; 115 classes are not connected at cosine ≥ 0.05; visualization based on the algorithm of Fruchterman & Reingold (1991).

At this level of fine-tuning, the German set appears as more integrated than the Chinese one. 560 of the 624 categories are used by the German set, as against 412 by the Chinese set; and 501 of the categories are related with a threshold of cosine ≥ 0.05 as against 349 in the Chinese case. However, in terms of structural properties, the two matrices (and the one for the global set) are again very flat. In the German case, for example, the first factor explains 0.32% of the common variance with an eigenvalue of 1.988, and 284 factors have an eigenvalue larger than one. Table 5 provides the network statistics in a format similar to that of Table 3 above.

| four digits;<br>*cosine* > 0.05 | *Global set*<br>(*N* = 624) | *Germany*<br>(*N* = 560) | *China*<br>(*N* = 412) |
|---|---|---|---|
| *Density* | 0.003 | 0.005 | 0.007 |
| *% Degree centralization* | 1.43 | 1.83 | 4.18 |
| *% Closeness centralization* | n.a. | n.a. | n.a. |
| *% Betweenness centralization* | 10.20 | 8.76 | 9.74 |
| *Clustering coefficient* | 0.345 | 0.215 | 0.356 |

**Table 5**: Network statistics of the cosine-normalized matrices for the German, Chinese, and global sets of patents classified at the four-digit level.

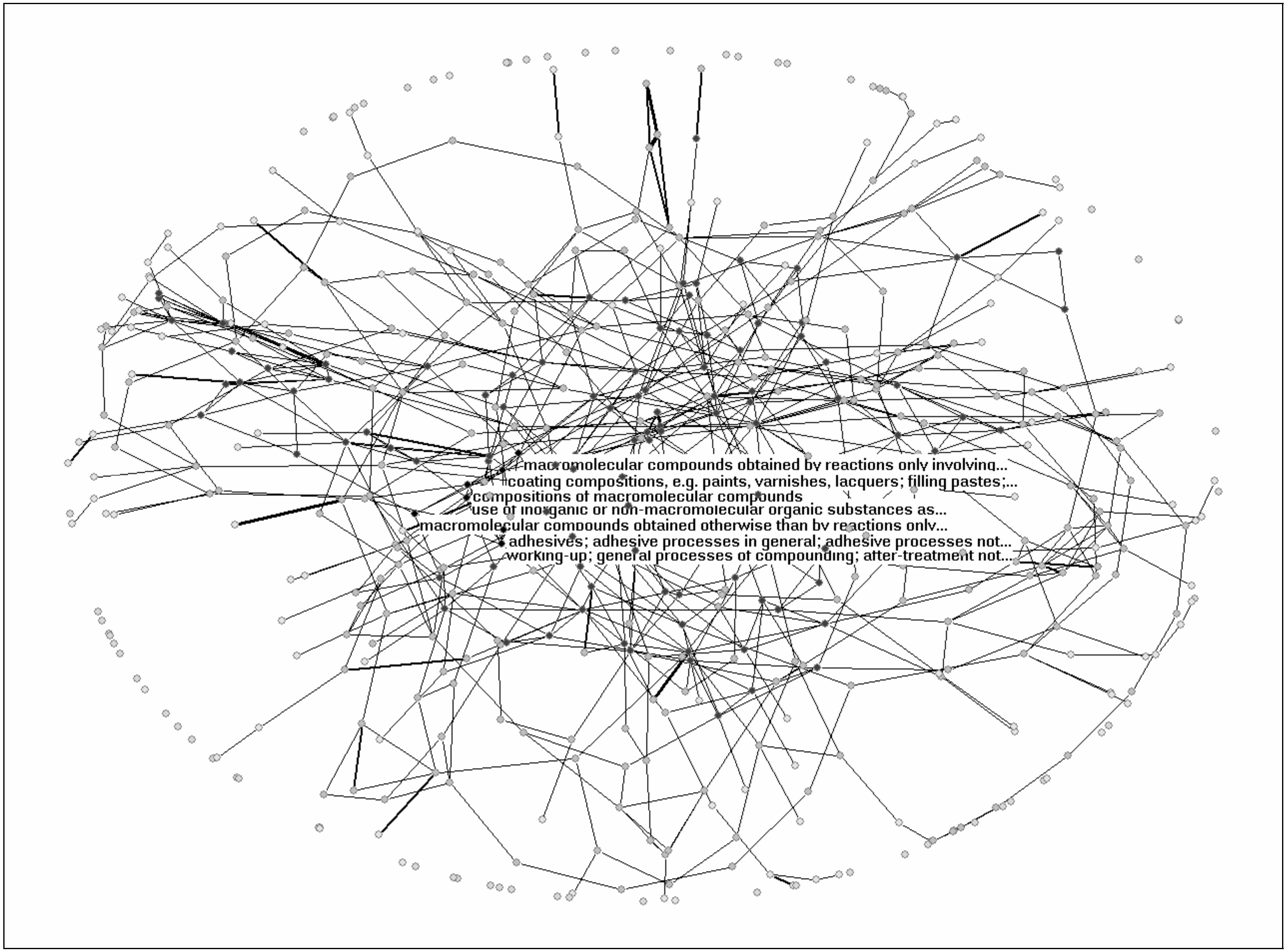
macromolecular compounds obtained by reactions only involving...
coating compositions, e.g. paints, varnishes, lacquers; filling pastes;...
compositions of macromolecular compounds
use of inorganic or non-macromolecular organic substances as...
macromolecular compounds obtained otherwise than by reactions only...
adhesives; adhesive processes in general; adhesive processes not...
working-up; general processes of compounding; after-treatment not...

**Figure 7**: 501 of the 560 patent classes are related at cosine ≥ 0.05 in the case of 17,095 patents with an inventor in Germany. The ($k$ = 10) core set is labeled; 2D-visualization based on the algorithm of Fruchterman & Reingold (1991).

This many categories cannot possibly be displayed meaningfully on a single screen, but the algorithms available in Pajek (and other visualization programs) enable us to filter out interesting subsets. In Figure 7, the set of nodes with the highest number of links among them ($k$ = 10) is displayed as an example. A similar exercise can be performed with the Chinese set.

Figure 8 shows a similar map for the much smaller Czech Republic. The structure of the core of this map is informative about the technological make-up of the country.

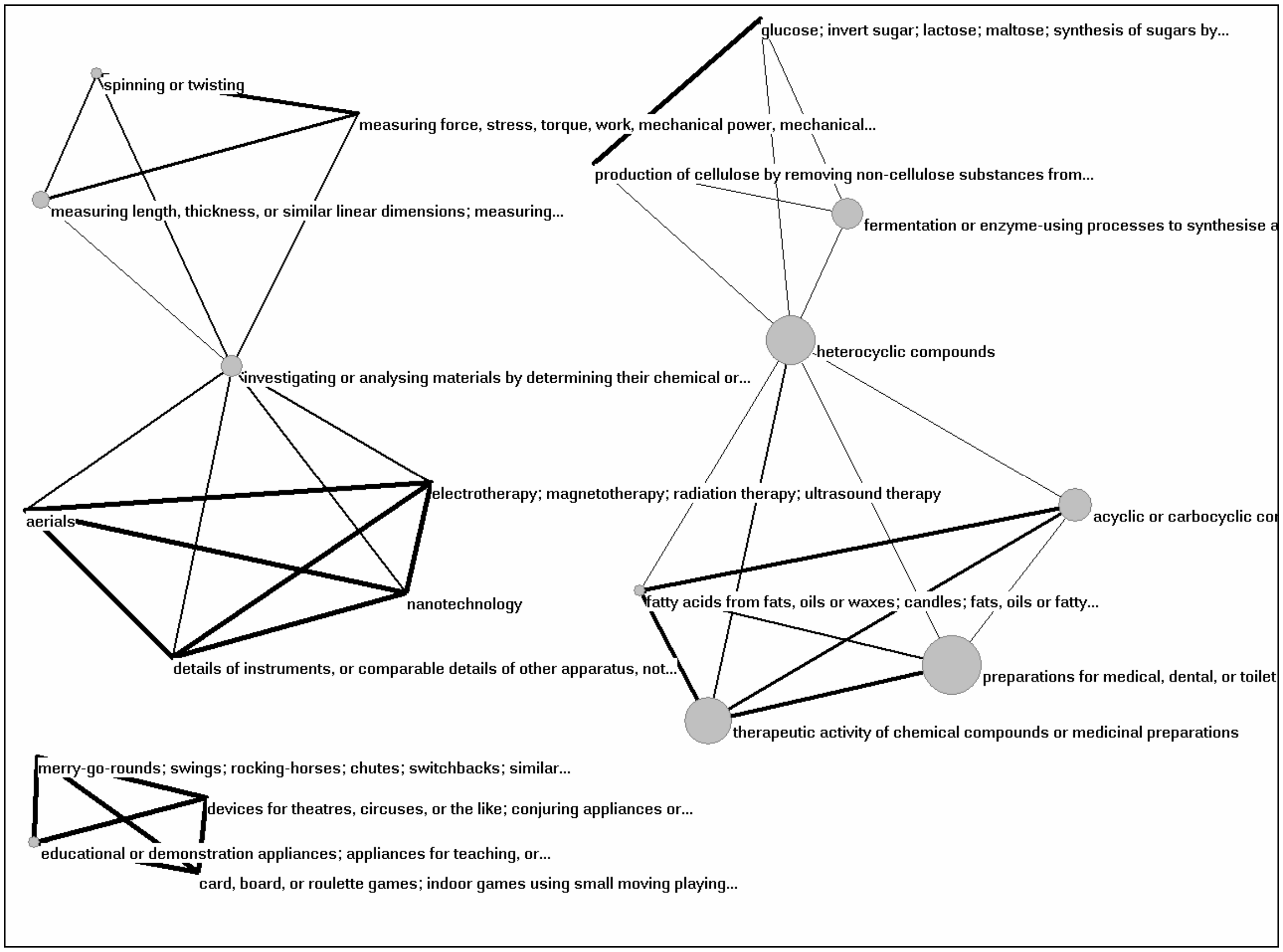

**Figure 8**: 20 patent classes in three clusters among the 113 which are listed for the Czech Republic; *N* of patents = 132; cosine ≥ 0.05. (Visualization based on the algorithm of Kamada & Kawai, 1989.)

In summary, and not surprisingly, the visualizations are more informative at the four-digit level: the countries are specific in terms of their portfolios. Table 6 compares Germany and China, analogously to Table 4, in terms of the percentage of betweenness centrality for the top 20 patent categories. Only three of the 20 categories match. Note that the added category "nanotechnology" (Y01N) ranks in the sixth place of this list for Germany.

| Germany | % | China | % |
|---|---|---|---|
| 1. layered products, i.e. products built-up of strata of flat or non-flat,... | 9.6 | 1. electric digital data processing | 10.6 |
| 2. spraying apparatus; atomising apparatus; nozzles | 8.9 | 2. separation | 9.8 |
| 3. cleaning in general; prevention of fouling in general | 5.7 | 3. semiconductor devices; electric solid state devices not otherwise... | 7.9 |
| 4. other working of metal; combined operations; universal machine tools | 5.1 | 4. investigating or analysing materials by determining their chemical or... | 7.8 |
| 5. mixing, e.g. dissolving, emulsifying, dispersing | 4.9 | 5. preparations for medical, dental, or toilet purposes | 5.1 |
| 6. nanotechnology | 4.7 | 6. containers for storage or transport of articles or materials, e.g.... | 5.0 |
| 7. lime; magnesia; slag; cements; compositions thereof, e.g. mortars,... | 4.5 | 7. layered products, i.e. products built-up of strata of flat or non-flat,... | 4.4 |
| 8. domestic plumbing installations for fresh water or waste water; sinks | 4.1 | 8. diagnosis; surgery; identification | 3.9 |
| 9. chemical or physical processes, e.g. catalysis, colloid chemistry;... | 4.1 | 9. kitchen equipment; coffee mills; spice mills; apparatus for making... | 3.7 |
| 10. gas-turbine plants; air intakes for jet-propulsion plants; controlling... | 3.8 | 10. devices for fastening or securing constructional elements or machine... | 3.6 |
| 11. macromolecular compounds obtained by reactions only involving... | 3.8 | 11. air-conditioning; air-humidification; ventilation; use of air currents... | 3.6 |
| 12. compounds of the metals beryllium, magnesium, aluminium, calcium,... | 3.8 | 12. pictorial communication, e.g. television | 3.1 |
| 13. processes for applying liquids or other fluent materials to surfaces,... | 3.7 | 13. methods or apparatus for sterilising materials or objects in general;... | 2.9 |
| 14. printing, duplicating, marking, or copying processes; colour printing | 3.4 | 14. processes or means, e.g. batteries, for the direct conversion of... | 2.8 |
| 15. making textile fabrics, e.g. from fibres or filamentary material;... | 3.4 | 15. household or table equipment | 2.5 |
| 16. abrasive or related blasting with particulate material | 3.2 | 16. foods, foodstuffs, or non-alcoholic beverages, not covered by... | 2.5 |
| 17. removal or treatment of combustion products or combustion residues;... | 3.1 | 17. printed circuits; casings or constructional details of electric... | 2.5 |
| 18. books; book covers; loose leaves; printed matter of special format or... | 3.1 | 18. macromolecular compounds obtained otherwise than by reactions only... | 2.4 |
| 19. launching, hauling-out, or dry-docking of vessels; life-saving in... | 3.1 | 19. chemical or physical processes, e.g. catalysis, colloid chemistry;... | 2.4 |
| 20. filling with liquids or semiliquids, or emptying, of bottles, jars,... | 3.0 | 20. working-up; general processes of compounding; after-treatment not... | 2.3 |

**Table 6**: top 20 classes at the four-digit level and the percentages of betweenness centrality for Germany and China, respectively.

### *5.3 Countries as units of analysis*

Are countries the appropriate unit of analysis? No, they are not. The data is generated bottom-up because of the national patent laws, but the PCT provides a protocol for streamlining the process and the IPC is developed as a classification at the level of the set. Some countries (e.g., Western European ones) may be very similar in terms of their patent portfolios.

Using this data, the common practice in evolutionary economics (e.g., Lundvall, 1992; Nelson, 1993) of using national systems as units for comparison can be investigated empirically (Foray & Lundvall, 1996; Leydesdorff, 2006a). When the patent data is aggregated at the level of countries, a matrix is generated with a single communality that explains 31.4% of the variance; factor 2 explains only 4.3% of the variance, and factor 3 less than 3.1%. The first communality corresponds with a large group of 78 countries (out of 126) which form a *k*-core in the network. If the threshold is set at cosine ≥ 0.05, only three countries are removed from the core set. At the level of cosine ≥ 0.5, 70 relatively advanced countries still form this dense core (Figure 9).

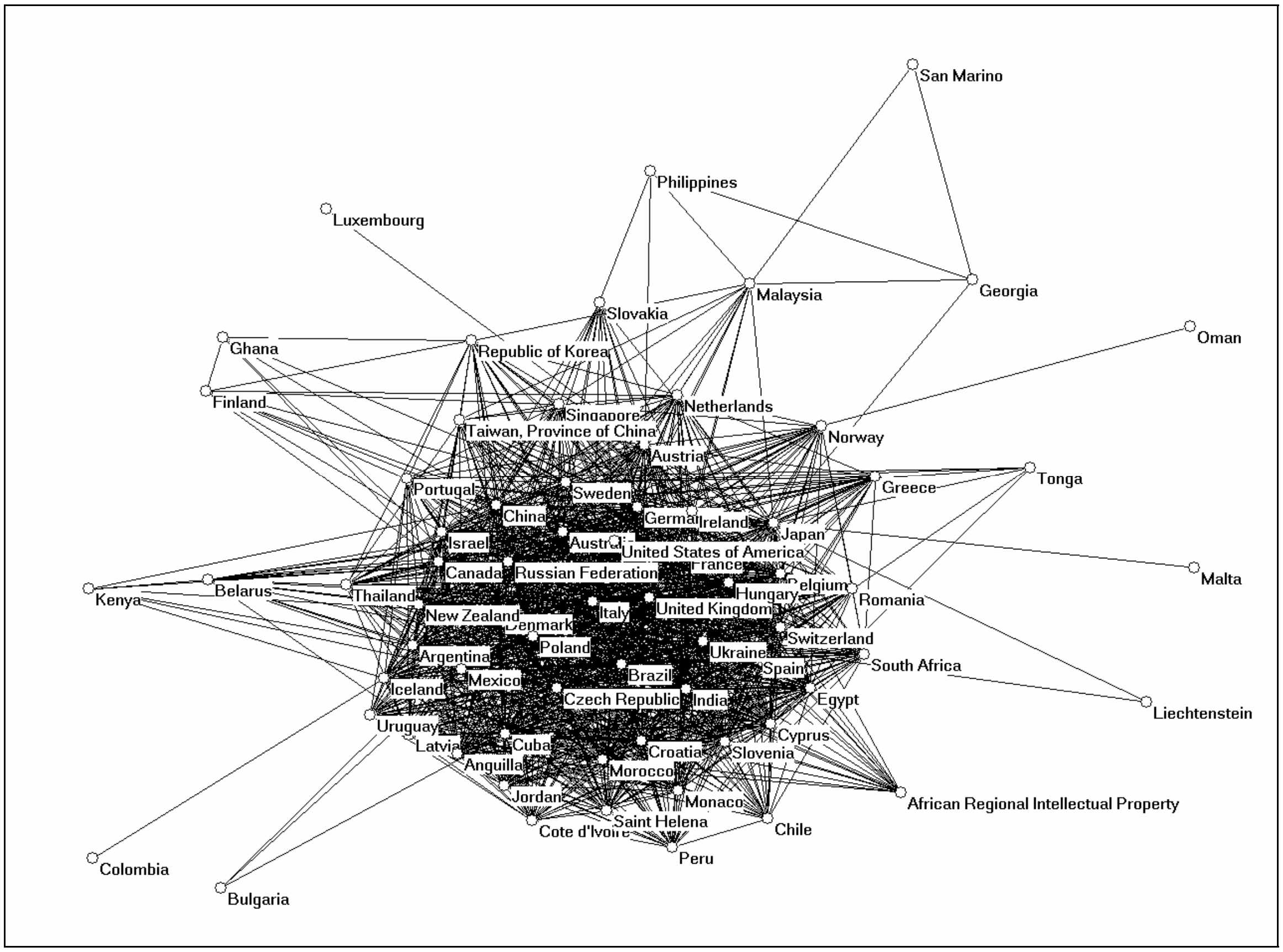

**Figure 9**: 70 countries form a dense core in terms of IPC classifications of their patent portfolios at the level of cosine ≥ 0.5.

Similarly, if one analyzes the four-digit codes used as variables in the underlying matrix, one obtains a first principal component that explains 65.4% of the common variance. 426 IPC categories are still connected at the level of cosine ≥ 0.5. In other words, most countries are very similar in terms of their patent portfolios and most patent classes are very similar in their distributions over the countries. In terms of patent classification, globalization has already taken place. The construction of the IPC at the global level may reinforce this abstraction from the national and institutional origins of the patent applications.

*5.4 Mapping emerging technologies*

The data-matrix used for the mapping of the four-digit classification codes at the world level (used for drawing Figure 6 above) enables us to select a specific class and to search for its relevant environment. (This file is brought online at http://www.leydesdorff.net/wipo06/world.zip.) This application of the instrument can be made policy relevant.

Using the recently added class for nanotechnology Y01N, for example, Figure 10 can be generated as its $k = 1$ environment at a threshold level of cosine $\geq 0.05$. Because the units of analysis are now again the patents themselves, the matrix is extremely sparse. The network layer spanned by the IPC is thin and has no pronounced structure.

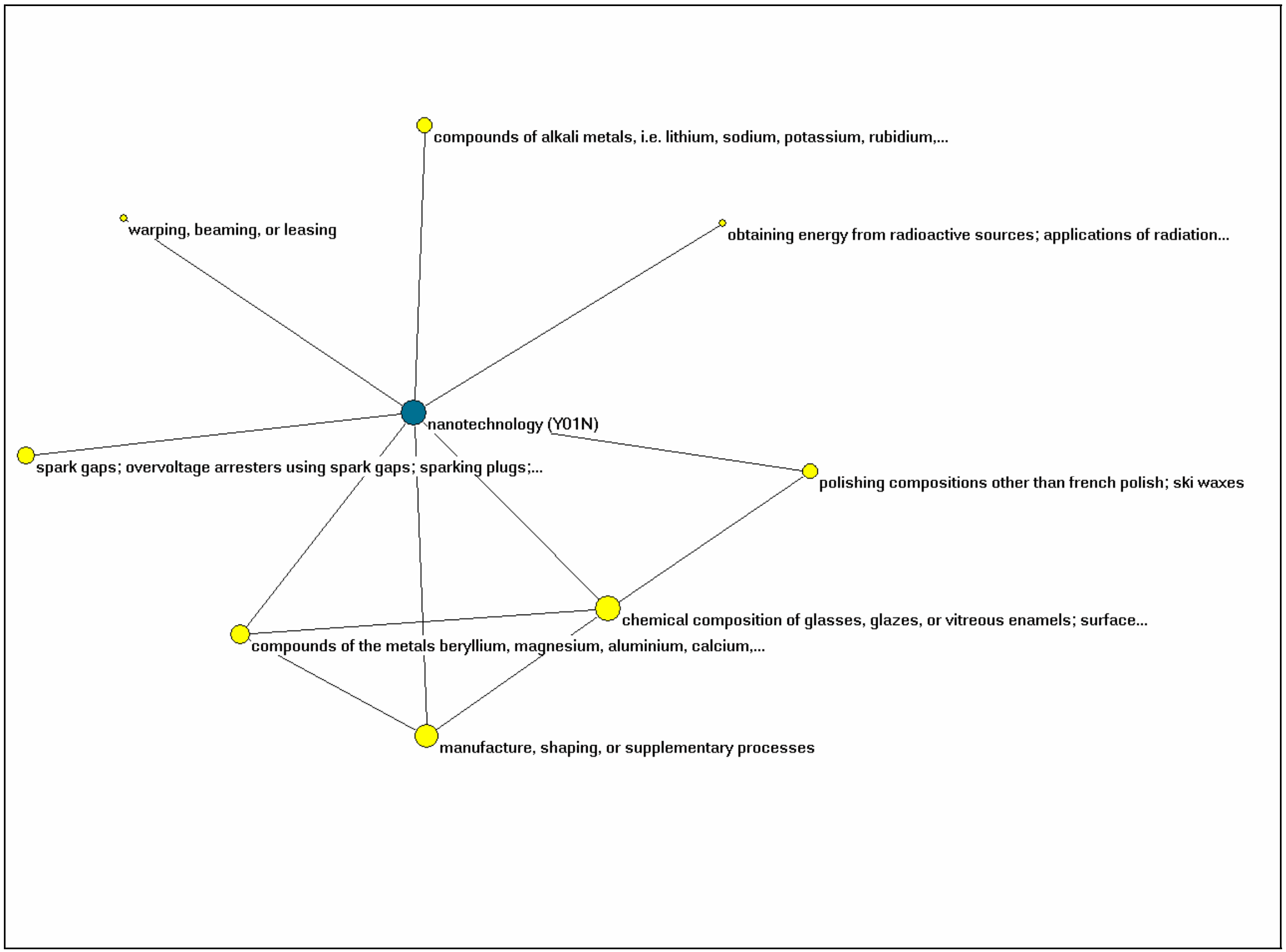

**Figure 10**: $k = 1$ neighborhood of class Y01N; $N = 762$; cosine $\geq 0.05$. (Visualization based on the algorithm of Kamada & Kawai, 1989.)

| | | | |
|---|---|---|---|
| USA | 330 | Austria | 4 |
| Japan | 120 | Australia | 4 |
| Germany | 88 | India | 4 |
| France | 46 | Denmark | 3 |
| United Kingdom | 34 | Greece | 3 |
| South Korea | 23 | Norway | 3 |
| Netherlands | 21 | Poland | 3 |
| Switzerland | 15 | Russia | 3 |
| Italy | 15 | Brazil | 2 |
| Canada | 13 | New Zealand | 2 |
| China | 11 | Turkey | 2 |
| Israel | 7 | Belarus | 1 |
| Sweden | 7 | Czech Republic | 1 |
| Belgium | 6 | Hong Kong | 1 |
| Spain | 6 | FYR Macedonia | 1 |
| Singapore | 6 | Mexico | 1 |
| Finland | 5 | Romania | 1 |
| Ireand | 5 | Taiwan | 1 |

| | | |
|---|---|---|
| | South Africa | 1 |

**Table 7**: The distribution of patents over (37) countries for the category “nanotechnology” (Y01N) using the WIPO dataset 2006 (762 patents; 799 addresses).

Table 7 lists the 37 countries which exhibit activity in this class using inventor addresses. Thus, the indicator can be made policy relevant. The relatively strong position of small countries like the Netherlands, Switzerland, and South Korea is again notable. Hullmann (2007, at p. 745) lists estimated public funding for these countries in 2004. The list of Table 7 correlates with $r = 0.97$; $p = 0.01$ ($N = 31$; 6 cases missing).

## 6. Conclusions

The major difference between the organization of scientific literature into journals which maintain and reproduce aggregated citation relations and the organization of patents into classes is a consequence of the role of the examiner. The examiner imposes additional citations and classifications for the purpose of *use*, while the journal structures emerge from the aggregated citation data in a self-organizing mode. From this perspective, patent classifications can be compared with the subject categories which the Institute of Scientific Information (of Thomson) attributes to journals (Leydesdorff & Rafols, in preparation). These categories are assigned by the ISI staff on the basis of a number of criteria, among which are the journal’s title, its citation patterns, etc. (McVeigh, *personal communication*, 9 March 2006). The classifications, however, match poorly with classifications derived from the database itself on the basis of analysis of the principal components of the networks generated by citations among them (Boyack *et al*., 2005; Leydesdorff, 2006b).

In the case of patents, classifications are attributed less arbitrarily. The patent offices make major investments in developing classification systems. Because of the depth of the classification system in terms of number of digits, one is able to zoom in or out of the system using a hierarchical structure. This is convenient for the human understanding, but it provides a thin layer for reflection on the underlying dynamics. The evolving database is captured in a dendogram. The associative relations *within* the dendogram can be made visible using co-classification analysis and provide us with a geometrical window on the complexity of the data. However, this is not an eigenstructure of the data, nor can one reveal the eigendynamics in the data by using these indicators. In other words, the status of these indicators is different from that of science indicators.

In the design of this study, the focus was on co-classifications as an alternative to co-citations because of the noted heterogeneity of functions of citations in the case of patent literature. In a follow-up study, a systematic comparison of co-classification patterns with citation patterns would be desirable using the USPTO set (because of the unification of citation formats within this set). One could then consider the classes as equivalents to journals and analyze the corresponding equivalent of an aggregated journal-journal citation matrix. This may work for statistical reasons despite the lack of retrievable structure in the co-classification patterns themselves (Leydesdorff & Rafols, in preparation).

The analysis taught us further that nations—which are the intuitive units of analysis because of national patent legislation—are not (or perhaps, no longer) the appropriate units of analysis for patent portfolios. The major structure at the global level seems the one between "haves" and "have-nots," or in other words, between countries included and those excluded from this technological realm. A majority of countries are included. The database is more apt for the analysis of how technologies are distributed among them in terms of the patent classifications. But even here, there seem to be no general rules of thumb, since the networks are sparse and can be expected therefore to remain highly sensitive to the parameters of the model, like the thresholds chosen, etc.

In summary, the results are a bit disappointing given the relatively well-organized dataset, and one should not expect better results from using more mixed sets like those currently under preparation by the OECD and the various patent offices. However useful from the perspective of management and policy making, "Tech Mining" (Porter & Cunningham, 2005) on the basis of institutionally composed databases can be expected to generate more fuzzy sets.

## 7. Discussion

The above conclusion may seem negative. However, this contribution is part of a discourse about the quality of various indicators for mapping. In this study, I analyzed (co)classifications because citations are a mixed bag in the case of patents more than in the case of scientific literature. In addition to citations and classifications, however, the patents as textual units also contain other textual elements, such as titles, abstracts, and full texts (Callon *et al.*, 1982, 1986; Mogoutov *et al.*, 2007). Words are less codified than citations (Leydesdorff, 1989), but in this case they may nevertheless be the best indicators of meaning that are available for the mapping (Leydesdorff & Hellsten, 2006). Let me illustrate this by using the patent portfolio of China.

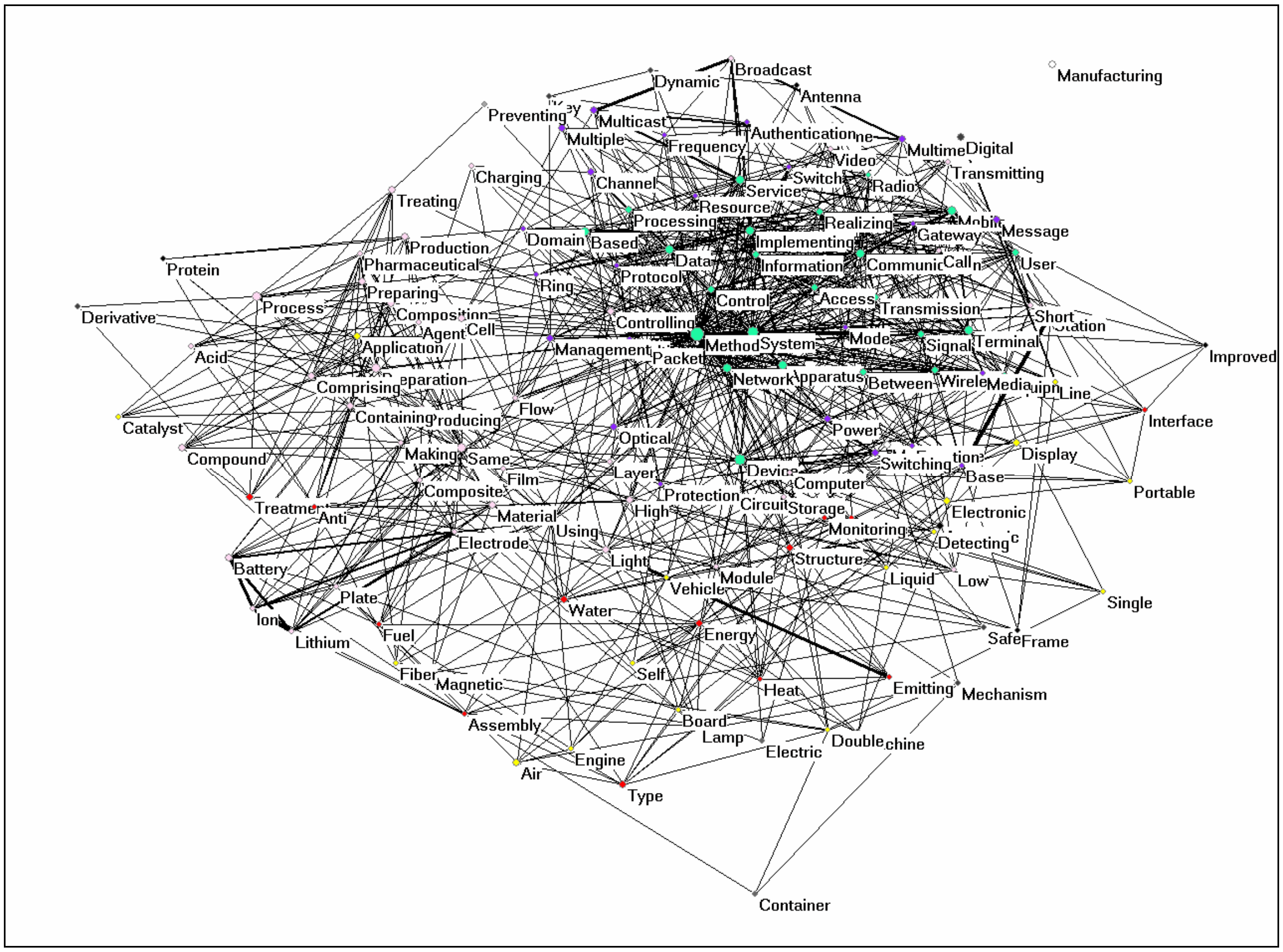

**Figure 11**: 139 words occurring more than twenty times in 3,084 titles of Chinese patents; cosine ≥ 0.05; visualization based on the algorithm of Kamada & Kawai (1989).

Figure 11 shows the cosine relations between 139 words that occur more than twenty times in 3,084 Chinese patents (used for the construction of Figure 5).[6] The picture reveals the focus on communication, computing, and networking in the Chinese patent portfolio. An analogous picture using the German patent portfolio (not shown here) exhibits the dominance of manufacturing. The contexts in which central words like "Methods," "Devices," and "Apparatuses" are provided with meaning are very different. For pragmatic reasons, these visualizations are limited to approximately 150 nodes on a single screen, but in terms of the statistics there are no limitations of this kind (Leydesdorff & Hellsten, 2006). Furthermore, the classifications enable us to delineate meaningful subsets whose contents can be analyzed further by using co-word (or citation!) analysis.

## Acknowledgements

I am grateful to Paola Criscuolo, Diana Lucio-Arias, Andrea Scharnhorst, Wilfred Dolfsma, David Gick, Martin Meyer, Thomas Gurney for advice and help in collecting the data. Three anonymous referees provided valuable comments and suggestions.

---

[6] The patents contain 4,0354 words which occur 18,1756 times. I used the stopword list of the USPTO available at http://ftp.uspto.gov/patft/help/stopword.htm.